# Resonant excitation of the domain wall oscillations by a parallel current under spin injection

R.G. Elliott

*University of Oxford, Department of Physics, Theoretical Physics,*
*1 Keble Road, Oxford OX1 3NP, United Kingdom*

E.M. Epshtein, Yu.V. Gulyaev, P.E. Zilberman[*]

*Institute of Radio Engineering and Electronics of the Russian Academy of Sciences, Fryazino Branch,*
*Vvedenskii sq. 1, Fryazino, Moscow District, 141190, Russia*

**Abstract**. A possibility is discussed of observing spin injection effect on the ferromagnet domain structure by means of resonant excitation of the domain wall oscillations by a spin-polarized ac injection current. The natural frequency of the domain wall oscillations in a thin ferromagnetic film with parallel anisotropy is calculated. Amplitude of the domain wall forced oscillations excited by the spin-polarized ac current is determined. Then effect of such oscillations on the current is considered and appearance of nonlinear phenomena such as rectification of the ac current and second harmonic generation is predicted.

1. Introduction

A possibility has been demonstrated [1, 2] of an influence of spin-polarized carriers on the domain structure of a ferromagnetic layer. S-d exchange interaction of the injected carriers with localized spins of the lattice decreases energy of the domains with magnetization parallel to the injector magnetization and increases energy of the domains with opposite magnetization. Therefore, the "parallel" domains become wider, while the "antiparallel" ones become narrower. Since the conductivities of the magnetic junctions with parallel and antiparallel relative orientation of the magnetic leads are different (the giant magnetoresistance effect [3]), such changes in the domain widths leads to changing the current through the junction.

In the present work, a possibility is discussed of observing spin injection effect on the ferromagnet domain structure by means of resonant excitation of the domain wall oscillations by a spin-polarized ac injection current. In Sec. 2, the natural frequency of the domain wall oscillations in a thin ferromagnetic film is calculated. Such a problem for the Faraday-type domain structure, in which the magnetization is perpendicular to the film developed surface, was solved in [4]. We consider the Cotton-type domain structure, in which the magnetization is parallel to the film developed surface [5]. In Sec. 3, amplitude of the domain wall forced oscillations excited by the spin-polarized ac current is determined. Then effect of such oscillations on the current value is considered and appearance of nonlinear phenomena such as rectification of the ac current and second harmonic generation is predicted.

---

[*] Corresponding author. Tel.: +7-095-526-9265; fax: +7-095-203-8414; e-mail: zil@ms.ire.rssi.ru

## 2. Domain wall free oscillations

We consider the same system as in [5], namely, a ferromagnetic thin film with parallel anisotropy and stripe domain structure. The $x$ axis is directed along the domain structure period, the $y$ axis is perpendicular to the film plane, and the $z$ axis is parallel to the magnetization vectors in the domains, i.e. is parallel to the anisotropy axis (Fig. 1). The domain walls parallel to $yz$ plane can oscillate along $x$ axis.

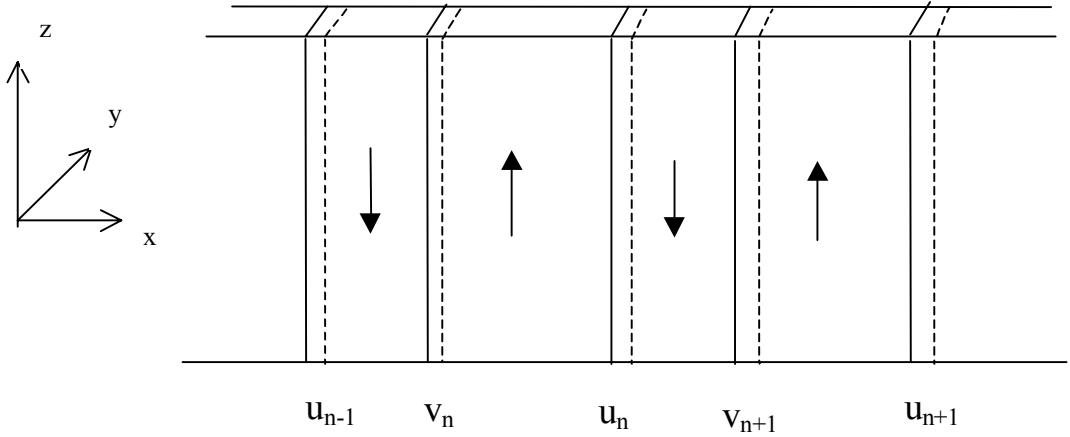

Fig. 1. The domain structure. The domain wall displacements are shown by dashed lines. The arrows show the magnetization directions in domains.

Calculations of the magnetostatic energy of a ferromagnetic thin film with parallel anisotropy and stripe domain structure were carried out in [5]. At $L_y \ll W_0 \ll L_z$ ($L_y$, $L_z$ are the film sizes, $W_0$ is spatial period of the domain structure) the magnetostatic energy per unit volume takes the following form:

$$E_M = \frac{2L_y}{L_z} M^2 \left\{ \left( \ln \frac{L_z}{L_y} + \frac{3}{2} \right) \left( \frac{W_-}{W_0} \right)^2 - \frac{4}{\pi^2} \sum_{k=1}^{\infty} \frac{1}{k^2} \left[ 1 - (-1)^k \cos \frac{k\pi W_-}{W_0} \right] \left( \ln \frac{k\pi L_y}{W_0} - \frac{3}{2} + \mathbf{C} \right) \right\}, \quad (1)$$

where $W_- = w_+ - w_-$, and $w_+$, $w_-$ are the widths of the domains with magnetization vector parallel and antiparallel to the $z$ axis positive direction, respectively, $M$ is the saturation magnetization including the equilibrium electron contribution, $\mathbf{C} = 0.5772\ldots$ is the Euler constant. The domain structure period is determined from the minimum condition of the sum of the magnetostatic energy and domain wall energy; it takes form

$$W_0 = \frac{L_z \gamma}{L_y M^2}, \tag{2}$$

where $\gamma$ is the domain wall energy per unit area.

Just as in [4], we assume that the (small) oscillations of the domain walls do not change their amount.

In absence of external magnetic field and injection current, the magnetostatic energy at given $W_0$ takes its minimum value at $W_- = 0$, i.e., with equal widths of the "parallel" and "antiparallel" domains.

If $W_-$ deviates from the equilibrium value, then a restoring force affects on the domain walls. Therefore, domain wall oscillations can occur. To estimate the frequency of such oscillations, we use the Döring model [6], which consider the domain wall as a massive particle under external forces. We take into account a possibility of nonuniform shifts of the domain walls and displacement waves propagating along $x$ axis.

It follows from Fig. 1 that the $u_n$ displacement of the right boundary of $n$th "parallel" domain is connected with $W_-$ by a relationship

$$W_- = 2u_n - v_n - v_{n+1}. \tag{3}$$

The $v_n$ displacement of the right boundary of $n$th "antiparallel" domain is described by analogous relationship

$$W_- = -(2v_n - u_n - u_{n-1}). \tag{4}$$

Since the number of the pairs of domain walls per unit length is $W_0^{-1}$, we find an equation of the domain wall motion from Eqs. (1), (3), and (4):

$$\frac{d^2 u_n}{dt^2} = -\frac{\partial E_M}{\partial u_n} = -\frac{1}{4}\omega_0^2 (2u_n - v_n - v_{n+1}), \tag{5}$$

$$\frac{d^2 v_n}{dt^2} = -\frac{\partial E_M}{\partial v_n} = -\frac{1}{4}\omega_0^2 (2v_n - u_n - u_{n-1}), \tag{6}$$

where

$$\omega_0^2 = \frac{32 L_y}{L_z} \frac{M^2}{mW_0} \left( \ln \frac{2L_z}{W_0} + \mathbf{C} \right), \tag{7}$$

$m$ is the domain wall mass per unit area.

Eqs. (5) and (6) are well known from the theory of oscillations of linear chains containing two kinds of particles [7]. Putting

$$u_n = u \exp[i(nqW_0 - \omega t)], \quad v_n = v \exp[i((n-\tfrac{1}{2})qW_0 - \omega t)], \tag{8}$$

we obtain dispersion relations

$$\omega_+(q) = \omega_0 \cos\frac{qW_0}{4}, \quad \omega_-(q) = \omega_0 \sin\frac{qW_0}{4}, \qquad (9)$$

corresponding to "optical" ($\omega_+$) and "acoustical" ($\omega_-$) modes of the domain wall oscillations (Fig. 2).

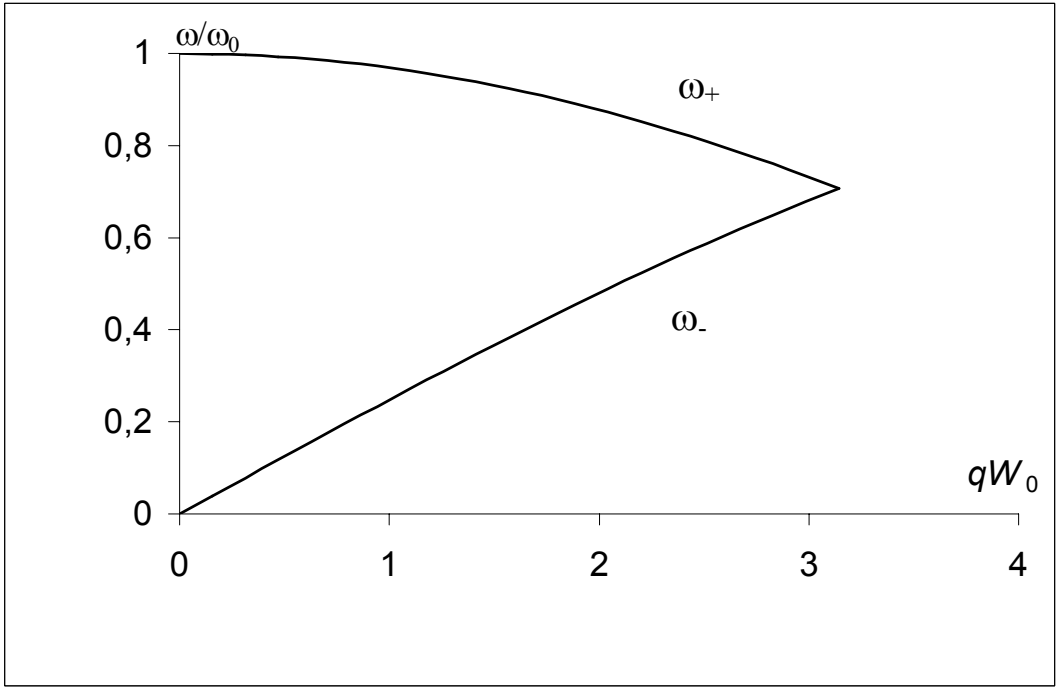

Fig. 2. Domain wall oscillation spectra.

Let us make some estimates. At $L_y = 50$ nm, $L_z = 50$ μm, $M = 0{,}1$ T, $m \sim 10^{-10}$ g/cm$^2$, and $\gamma \sim 10^{-7}$ J/cm$^2$, we have $\omega_0/2\pi \sim 100$ MHz.

### 3. Forced oscillations of the domain walls

Consider a system consisted of a spin injector in $y < 0$ semispace with magnetization in the positive direction of $z$ axis, the film with domain structure in $0 < y < L_y$ range, and a nonmagnetic lead in $y > L_y$ range. As mentioned in Sec. 1, the spin injection into the film and s-d exchange interaction of the injected spins with localized lattice spins lead to widening "parallel" domains and narrowing "antiparallel" ones.

The s-d exchange interaction energy density is [8]

$$E_{s-d} = \frac{1}{2}\left[E_{s-d}^{(p)}\left(1+\frac{W_-}{W_0}\right) + E_{s-d}^{(a)}\left(1-\frac{W_-}{W_0}\right)\right], \qquad (10)$$

$$E_{s-d}^{(p,a)} = \mp\frac{\alpha\mu_B M\tau[Q_1\sinh\lambda + \nu(Q_1 \mp Q_2)(\cosh\lambda-1)]}{eL_y(\sinh\lambda + \nu\cosh\lambda)} j_{p,a} \equiv k_{p,a} j_{p,a}, \qquad (11)$$

where $\alpha$ is a dimensionless s-d exchange interaction constant, $\mu_B$ is the Bohr magneton, $\tau$ is the the injected spin relaxation time, $\lambda = L_y/l$, $l$ is the injected spin diffusion length, $\nu = (\sigma_0/\sigma_F)(l_F/l_0)/(1-Q_2^2)$, $\sigma_F$ and $\sigma_0$ are conductivities of the ferromagnetic film and the nonmagnetic lead, respectively, $Q_1$, $Q_2$ are the current spin polarizations in the injector and ferromagnetic film, respectively, and $j_p$, $j_a$ are the injector current densities in "parallel" and "antiparallel" domains, respectively; the top signs in (11) refer to "parallel" domains and bottom ones to "antiparallel" domains.

It follows from (5), (6) and (10) that an additive term $\mp\frac{1}{m}\left(E_{s-d}^{(p)} - E_{s-d}^{(a)}\right)$ having a sense of an external force appears in the right hand side of the Eqs. (5) and (6) in presence of the s-d exchange interaction. Hence, a possibility appears of exciting forced oscillations of the domain walls by ac injection current. The excitation has resonant character when the ac current frequency coincides with any of natural frequencies.

In particular, uniform ($q = 0$) "optical" small oscillations of the domain walls, changing relative width of the "parallel" and "antiparallel" domains under ac electric field of $\Omega$ frequency applied to the junction, are described by the equation of a harmonic oscillator with ac external force

$$\frac{d^2 W_-}{dt^2} + \beta\frac{dW_-}{dt} + \omega_0^2 W_- = A\cos\Omega t, \qquad (12)$$

where $\beta$ is a phenomenological damping constant, $A = -\frac{4(k_p j_{p0} - k_a j_{a0})}{mW_0}$, $j_{p0}$, $j_{a0}$ are amplitudes of the current densities.

It follows from Eq. (12)

$$W_-(t) = W_-^{(0)}\cos(\Omega t + \varphi), \qquad (13)$$

$$W_-^{(0)} = \frac{A}{\sqrt{(\Omega^2 - \omega_0^2)^2 + \beta^2\Omega^2}}, \quad \varphi = \arctan\frac{\beta\Omega}{\Omega^2 - \omega_0^2}. \qquad (14)$$

At $\beta \ll \omega_0$, $\Omega \approx \omega_0$ the excitation has resonant character.

The average current density through the junction is

$$\bar{j} = (j_p w_+ + j_a w_-)/W_0 = \tfrac{1}{2}(j_p + j_a) + (j_p - j_a)(W_-(t)/W_0)$$
$$= \tfrac{1}{2}(j_{p0} + j_{a0})\cos\Omega t + \tfrac{1}{2}(j_{p0} - j_{a0})(W_-^{(0)}/W_+)[\cos\varphi + \cos(2\Omega t + \varphi)]. \qquad (15)$$

It follows from Eq. (15) that oscillations of domain walls excited by ac spin injection current are accompanied with appearance of a dc component (the synchronous detection effect) and the current second harmonic. The values of such components are proportional to difference between the junction conductivities under parallel and antiparallel relative orientation of the magnetic layers, i.e., proportional to the junction giant magnetoresistance. The dc component and the second harmonic depend on the frequency of the applied ac voltage by a resonant way.

The effect considered can be used for experimental observing influence of the spin injection on the ferromagnetic domain structure.

The work was supported by International Science and Technology Center (Grant No. 1522) and Russian Foundation of Basic Research (Grant No. 03-02-17540).